\documentclass[journal,twoside]{IEEEtran}

\usepackage{makeidx,epsfig}
\usepackage{setspace,graphicx,multirow}

\usepackage{amsfonts,latexsym,amssymb,amsthm}

\usepackage{graphicx}
\usepackage{epstopdf}

\usepackage{caption3} 
\DeclareCaptionOption{parskip}[]{} 
\usepackage[small]{caption}

\usepackage{subcaption}

\usepackage{array}

\usepackage[cmex10]{amsmath}

\usepackage{url}

\usepackage{multirow}
\usepackage{hhline}

\usepackage{amsmath}
\usepackage{algorithm}
\usepackage[]{algpseudocode}
\usepackage{varwidth}

\makeatletter
\def\BState{\State\hskip-\ALG@thistlm}
\makeatother

\usepackage{algcompatible}

\usepackage{fixltx2e}

\usepackage{color}

\begin{document}
\bibliographystyle{IEEEtran}

\title{Placement Optimization of UAV-Mounted Mobile Base Stations}

\author{
Jiangbin~Lyu,~\textit{Member,~IEEE},
        Yong~Zeng,~\textit{Member,~IEEE},
        Rui~Zhang,~\textit{Fellow,~IEEE}
        and~Teng~Joon~Lim,~\textit{Fellow,~IEEE}%
\thanks{The authors are with the Department of Electrical and Computer Engineering, National University of Singapore (email: \{elelujb, elezeng, elezhang, eleltj\}@nus.edu.sg).}%
\vspace{-3ex}}

\maketitle

\begin{abstract}
In terrestrial communication networks without fixed infrastructure, unmanned aerial vehicle (UAV)-mounted mobile base stations (MBSs) provide an efficient solution to achieve wireless connectivity. This letter aims to minimize the number of MBSs needed to provide wireless coverage for a group of distributed ground terminals (GTs), ensuring that each GT is within the communication range of at least one MBS.
We propose a polynomial-time algorithm with successive MBS placement, where the MBSs are placed sequentially starting on the area perimeter of the uncovered GTs along a spiral path towards the center, until all GTs are covered. Each MBS is placed to cover as many uncovered GTs as possible, with higher priority given to the GTs on the boundary to reduce the occurrence of outlier GTs that each may require one dedicated MBS for its coverage.
Numerical results show that the proposed algorithm performs favorably compared to other schemes in terms of the total number of required MBSs and/or time complexity.

\end{abstract}


\vspace{-1ex}
\section{Introduction}

With their maneuverability and increasing affordability, unmanned aerial vehicles (UAVs) have many potential applications in wireless communication systems \cite{ZengUAVmag}.
In particular, UAV-mounted mobile base stations (MBSs) can be deployed to provide wireless connectivity in areas without infrastructure coverage such as battlefields or disaster scenes.
Unlike terrestrial base stations (BSs), even those mounted on ground vehicles, UAV-mounted MBSs can be deployed in any location and move along any trajectory constrained only by their aeronautical characteristics, in order to cover the ground terminals (GTs) in a given area based on their known locations.
When the UAV-GT channels are dominated by line-of-sight (LOS) links, the authors in \cite{PlacementTwoTier} use a K-means clustering algorithm to partition the GTs to be served by $p$ UAVs, while each UAV has a capacity constraint and the unsupported GTs are served by the fixed ground BSs.
The authors in \cite{3Dplacement} adopt a probabilistic LOS channel model and study the 3-dimensional (3D) placement of a single aerial BS to offload as many GTs as possible from the ground BS.

\begin{figure}
\centering
\includegraphics[width=0.8\linewidth,  trim=0 0 0 0,clip]{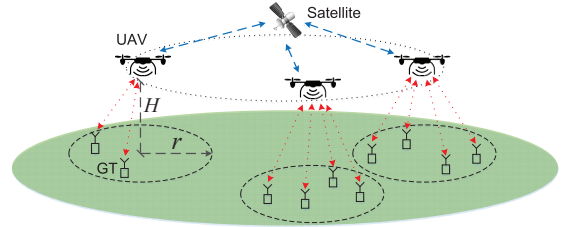} 
\caption{A wireless communication system with UAV-mounted MBSs\vspace{-3ex}} \label{SpiralScheme}
\end{figure}

In this letter, we assume that the GT locations are known and the UAVs are flying at a fixed altitude $H$, while the UAV-GT channels are dominated by LOS links whose channel quality mainly depends on the UAV-GT distance. 
We consider the scenario where no ground BSs are available and the UAV-mounted MBSs are backhaul-connected via satellite links, while each MBS has an equivalent coverage radius of $r$ projected on the ground, as shown in Fig. \ref{SpiralScheme}.
We thereby focus on the MBS placement problem to provide wireless coverage for all GTs in a given area. 
This can be formulated as the Geometric Disk Cover (GDC) problem \cite{TONbackbone},
whose objective is to cover a set of $K$ nodes (GTs) in a region with the minimum number of disks of given radius $r$.
The GDC problem can be optimally solved by the core-sets method \cite{CoresetEgypt} whose theoretical bounds on the running time are exponential in $K$.
Since the GDC problem is NP-hard in general,
a strip-cover-with-disks algorithm was proposed in \cite{TONbackbone},
which divides the plane into equal-width strips and solves the problem locally over the GTs within each strip. 
The computational complexity is reduced thanks to this strip-based partitioning which, however, may lead to significant performance loss since the GTs in different strips are independently considered though certain GTs in adjacent strips could in fact be covered by the same MBS.

This letter proposes a new MBS placement algorithm
by placing the MBSs sequentially,
starting from the perimeter of the area boundary in an inward spiral manner until all GTs are covered.
In the proposed spiral placement algorithm,
each MBS is first positioned to cover at least one uncovered GT near the area boundary, and then its position is adjusted inwards toward the area center to cover as many additional uncovered GTs as possible.
This localized strategy
has low complexity and does not partition the coverage area into independent regions,
hence overcoming the limitations of the strip-based algorithm. 
Our proposed algorithm has a polynomial-time complexity $O(K^3)$ in the worst case, which is comparable to the strip-based algorithm but much lower than the core-sets method.
Numerical results show that
for small networks requiring only a few MBSs, where the theoretical minimum can be found by the core-sets method, 
the proposed algorithm provides the near-optimal performance in terms of the number of required MBSs.
Moreover, the proposed algorithm also outperforms other heuristic schemes in terms of the number of required MBSs and time complexity on average for networks of different sizes, including the strip-based algorithm, the K-means clustering algorithm, and the random placement algorithm.
Note that the proposed algorithm can be considered as a new approach to solve the GDC problem in general and thus can be used in other pertinent applications.

\section{System Model and Problem Formulation}\label{SysModel}

We consider a wireless system with $K$ GTs, which are denoted by the set $\mathcal{K}=\{1,2,\cdots,K\}$ and at known locations given by $\{\textbf{w}_k\}_{k\in\mathcal{K}}$, where $\textbf{w}_k\in \mathbb{R}^{2\times 1}$ represents the two-dimensional (2D) coordinates of the $k$-th GT on the horizontal plane (ground). 
Assume that the UAV-GT communication channels are dominated by LOS links. 
Though simplified, the LOS model 
offers a good approximation for practical UAV-GT channels, and enables us to investigate the coverage problem in this letter. Other practical issues such as multi-access and UAV-GT association can be considered separately \cite{CyclicalLyu}.
We assume that the transmit power is fixed and the minimum required signal-to-noise ratio (SNR) at the receiver for reliable communications is given. Under the LOS model, the UAV-GT channel power gain follows the free-space path-loss model, which is determined by the UAV-GT link distance.
Assume that the UAVs are flying at a given altitude $H$ and their maximum coverage radius projected on the ground plane corresponding to the SNR threshold is specified by $r$, as shown in Fig. \ref{SpiralScheme}.

For cost minimization, we aim to deploy the minimum number of MBSs (UAVs) so that each GT is served by at least one MBS within its communication radius $r$.
Note that this does not preclude the possibility that some GTs may be covered by more than one MBSs. In such scenarios, the inter-cell interference issue needs to be addressed by, e.g., proper channel assignment and power control after deploying the MBSs, which is out of the scope of this letter.
Denoting by $\mathcal M=\{1,...,M\}$ the set of MBSs to be deployed, the problem can be formulated as follows.
\begin{align}
\mathrm{(P1)}:
\begin{cases}
\underset{\{\textbf{u}_m\}_{m\in\mathcal{M}}}{\min} & \ |\mathcal{M}|   \notag \\
\text{s.t.}  & \ \underset{m\in \mathcal{M}}{\min} \|\mathbf w_k - \mathbf u_m\|\leq r, \ \forall k\in \mathcal{K},
\end{cases}
\end{align}%
where $|\mathcal{M}|=M$ denotes the cardinality of the set $\mathcal{M}$, $\textbf{u}_m\in \mathbb{R}^{2\times 1}$ denotes the horizontal coordinates of MBS $m$, and the Euclidean norm $\|\mathbf w_k - \mathbf u_m\|$ is the distance between GT $k$ and MBS $m$ projected on the ground plane.

(P1) is also known as the GDC problem \cite{TONbackbone}, which is NP hard in general. 
The GDC problem is also related to the $p$-center problem \cite{KcenterExact1998}, which aims to locate $p$ centers (MBS locations) of the smallest disks to cover all $K$ nodes (GTs), given by
\begin{align}
\mathrm{(P2)}:
\begin{cases}
\underset{\{\textbf{u}_m\}_{m=1}^p}{\min} & \ \rho   \notag \\
\text{s.t.}  & \ \underset{m=1,\cdots,p}{\min} \|\mathbf w_k - \mathbf u_m\|\leq \rho, \ \forall k\in \mathcal{K},
\end{cases}
\end{align}%
whose optimal value $\rho^*$ is the smallest radius of the $p$ disks required to cover all $K$ GTs.
If $\rho^*\leq r$, then all GTs can be covered by the $p$ MBSs in (P1) and $M_{\min}\leq p$, where $M_{\min}$ denotes the optimal value of (P1).
The GDC problem (P1) can thus be converted into a series of $p$-center problems with increasing $p$ values, until the smallest number of MBSs required to cover all GTs is found. 
Unfortunately, (P2) is in general difficult to solve optimally due to the non-convex constraint $\underset{m=1,\cdots,p}{\min} \|\mathbf w_k - \mathbf u_m\|\leq \rho$, $\forall k\in \mathcal K$, whose left-hand side is the minimum of convex functions and hence is non-convex.
In fact, 
the $p$-center problem is also NP-hard, whose optimal solution 
requires computational complexity of $O(p^K)$ using brute force search \cite{KcenterComplexity}, which is infeasible even for moderate values of $p$ and $K$. 
Recent progress is based on the exploitation of a small subset of GTs called core-sets \cite{CoresetEgypt}.
A branch-and-bound algorithm to traverse the partitions of possible core-sets using depth-first strategy is given in \cite{CoresetEgypt}, which can find the optimal solution to the $p$-center problem for small values of $p$ ($p\leq 8$), 
although the worst-case complexity is still $O(p^K)$. 


\section{Spiral MBS Placement Algorithm}\label{SectionSDC}

In this section, we propose an efficient heuristic algorithm to solve (P1) approximately based on successive MBS placement. The main idea is to place the MBSs sequentially
along the area perimeter, which is defined as the path connecting the extreme points (referred to as the boundary GTs) of the convex hull of all uncovered GTs.
Each MBS $m$ is guaranteed to cover at least one boundary GT $k_0$, and those GTs at a distance of more than $2r$ away from $k_0$ are removed from consideration, since they cannot be jointly covered with $k_0$ by the same MBS $m$.
Since $k_0$ is at the boundary, MBS $m$ will be placed inwards toward the area center to cover as many uncovered GTs as possible, with higher priority given to the GTs on the boundary to reduce the occurrence of outlier GTs that each may require one dedicated MBS for its coverage.
After MBS $m$ is placed, the area perimeter of the remaining uncovered GTs shrinks at the local region near $k_0$.
The above process repeats to place the next MBS $m+1$ counterclockwisely next to MBS $m$, and the area perimeter gradually shrinks until all GTs are covered. As a result, the connecting line of the placed MBSs looks like a spiral which starts from the area boundary and counterclockwisely revolves inwards toward the area center.
We therefore name our proposed algorithm as the \textit{spiral MBS placement} algorithm, which is summarized in Algorithm \ref{AlgSpiral}.
\begin{algorithm}[H]\caption{Spiral MBS Placement Algorithm}\label{AlgSpiral}
\begin{small}
\textbf{Input:} GT set $\mathcal K$, with known locations $\{\mathbf w_k\}_{k\in\mathcal{K}}$.\\
\textbf{Output:} MBS set $\mathcal M$, with optimized locations $\{\mathbf u_m\}_{m\in\mathcal{M}}$.\\
\textbf{Initialization:} Uncovered GT set $\mathcal K_U\leftarrow \mathcal K$; $\mathcal{M}=\emptyset; m=1$.
\begin{algorithmic}[1]
\WHILE{$\mathcal K_U \neq \emptyset$}
\STATE Find boundary GT set $\mathcal K_{U,bo}\subseteq\mathcal K_U$ and list them in counterclockwise order. Update inner GT set $\mathcal K_{U,in}\leftarrow\mathcal K_U\setminus \mathcal K_{U,bo}$. If $m=1$, randomly pick a GT $k_0\in \mathcal K_{U,bo}$.
\STATE Refine MBS location $\mathbf{u}$ to cover $k_0$ and as many boundary GTs as possible, by calling $[\mathbf{u},\mathcal P_{prio}]$ = \textbf{LocalCover}($\mathbf{w}_{k_0}$, $\{k_0\}$, $\mathcal K_{U,bo}\setminus \{k_0\}$). Let $\mathcal K_{new,bo}\leftarrow \mathcal P_{prio}$. 
\STATE Refine MBS location $\mathbf{u}$ to cover $\mathcal K_{new,bo}$ and as many inner GTs as possible, by calling $[\mathbf{u},\mathcal P_{prio}]$ = \textbf{LocalCover}($\mathbf{u}$, $\mathcal K_{new,bo}$, $\mathcal K_{U,in}$). Let $\mathbf u_m=\mathbf u$, $\mathcal K_{new}\leftarrow \mathcal P_{prio}$. 
\STATE $\mathcal M\leftarrow\mathcal M\cup\{m\}$, $\mathcal K_{U}\leftarrow\mathcal K_{U}\setminus\mathcal K_{new}$, $m\leftarrow m+1$.
\STATE From $\mathcal K_{U,bo}\setminus \mathcal K_{new,bo}$, pick the first uncovered boundary GT $k_0'$ counterclockwisely next to $k_0$. Let $k_0\leftarrow k_0'$.
\ENDWHILE
\end{algorithmic}
\end{small}
\end{algorithm}

\begin{figure}
\centering
\includegraphics[width=0.85\linewidth,  trim=0 0 0 0,clip]{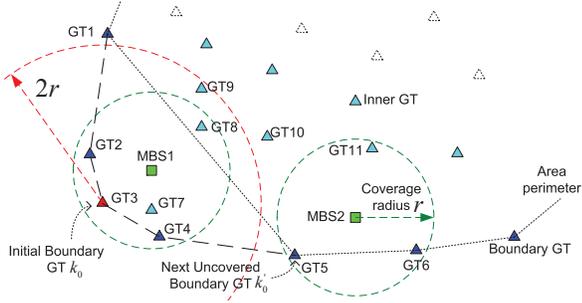} 
\caption{Illustration of the spiral algorithm\vspace{-3ex}} \label{SpiralExample}
\end{figure}

We use the example in Fig. \ref{SpiralExample} to illustrate the notations and the main steps of our spiral algorithm.
Denote by $\mathcal{K}_U\subseteq\mathcal{K}$ the subset of uncovered GTs, which is initialized to $\mathcal{K}$ at the beginning of Algorithm \ref{AlgSpiral}.
$\mathcal{K}_U$ is partitioned into the inner GT subset $\mathcal K_{U,in}$ and the boundary GT subset $\mathcal K_{U,bo}$, where the boundary GTs can be listed in counterclockwise order as $\mathcal K_{U,bo}=\{1,2,3,4,5,6,\cdots\}$ initially (dark blue triangles), and $\mathcal K_{U,in}=\mathcal K_U\setminus \mathcal K_{U,bo}$ (light blue triangles).
The path connecting these boundary GTs is referred to as the area perimeter of the uncovered GTs, as shown in Fig. \ref{SpiralExample}.
We use the convex hull to define the boundary GTs, whereas other boundary definitions \cite{CharacteristicHull} can also be used which produce similar results.

We give higher priority to the boundary GTs in the way that a certain subset of boundary GTs are guaranteed to be covered by each newly placed MBS.
To place the first MBS, we randomly choose a boundary GT $k_0$ which is guaranteed to be covered, e.g., GT 3 at the lower left corner denoted by a red triangle (step 2 in Algorithm \ref{AlgSpiral}).
Then we refine the MBS location $\textbf{u}$ to cover $k_0$ and as many boundary GTs as possible (step 3). 
In this case, the boundary GTs 2 and 4 can be covered, and hence are added into the prioritized set $\mathcal P_{prio}=\{2,3,4\}$ which is guaranteed to be covered first. Then we proceed to cover GTs from $\mathcal P_{prio}$ and as many inner GTs as possible (step 4). In this case, the inner GTs 7 and 8 can be covered.
The final location of the first MBS is denoted by a green square, which is the center of the covering disk of radius $r$, denoted by a dashed green circle.
After placing the first MBS, the area perimeter shrinks at the local region near GT $k_0$, with GT 1 directly connected to GT 5 in this case.
To place the next MBS, we pick the first uncovered boundary GT $k_0'$ counterclockwisely next to $k_0$, which in this case is GT 5, and update $k_0\leftarrow k_0'$ (step 6).
Then the above steps are repeated to place the second MBS which covers GTs 5, 6 and 11.
The above process repeats until all GTs are covered.

Note that we have used a \textbf{LocalCover} procedure in steps 3 and 4 of Algorithm \ref{AlgSpiral}, which refines the new MBS location $\textbf{u}$ to guarantee to cover GTs from the given prioritized set $\mathcal P_{prio}$ (e.g., the initial boundary GT $k_0$), and then to cover as many GTs as possible from the secondary set of GTs (e.g., uncovered inner GTs), denoted as $\mathcal P_{sec}$.
Mathematically, this can be formulated as the following optimization problem.
\begin{align}
\mathrm{(P3)}:
\begin{cases}
\underset{\mathbf u, \mathcal K_{new}}{\max} & \ |\mathcal K_{new}|   \notag \\
\text{s.t.}  & \ \|\mathbf u-\mathbf w_k\| \leq r, \ \forall k\in \mathcal K_{new}\cup \mathcal P_{prio},\\
& \ \mathcal K_{new} \subseteq \mathcal P_{sec},
\end{cases}
\end{align}%
where $\mathbf u$ denotes the location of the new MBS to be placed, $\mathcal K_{new} \subseteq \mathcal P_{sec}$ denotes the set of GTs newly covered by this new MBS. Note that the first constraint in (P3) ensures that all GTs in $\mathcal K_{new}$ and $\mathcal P_{prio}$ are covered by this new MBS. 
(P3) is a combinatorial optimization problem, which in general requires exhaustive search over all $2^{|\mathcal P_{sec}|}$ subsets of $\mathcal P_{sec}$ in order to obtain the optimal solution, which is prohibitive even for moderately large systems. 
Therefore, we propose a \textbf{LocalCover} procedure with possibly sub-optimal solutions to (P3) for low-complexity implementation, as summarized in Algorithm \ref{AlgLocal}.
\begin{algorithm}[H]\caption{\textbf{LocalCover} Procedure}\label{AlgLocal}
\begin{small}
\textbf{Procedure} $[\mathbf{u},\mathcal P_{prio}]$ = \textbf{LocalCover}($\mathbf{u}$, $\mathcal P_{prio}$, $\mathcal P_{sec}$)
\begin{algorithmic}[1]
\WHILE{$\mathcal P_{sec} \neq \emptyset$}
\STATE Update $\mathcal P_{sec}$ by excluding GTs more than $2r$ away from any GT in $\mathcal P_{prio}$. Update $\mathcal P_{prio}$ ($\mathcal P_{sec}$) by including (excluding) GTs within distance $r$ to $\mathbf u$.
\STATE Find GT $k_1\in\mathcal P_{sec}$ with shortest distance to $\mathbf u$. Add (remove) $k_1$ to (from) $\mathcal P_{prio}$ ($\mathcal P_{sec}$) if it can be covered by refining $\mathbf u$ via solving the 1-center problem. Stop otherwise.
\ENDWHILE
\end{algorithmic}
\end{small}
\end{algorithm}

We continue to use the example in Fig. \ref{SpiralExample} to illustrate Algorithm \ref{AlgLocal}. 
Firstly, for any given $\mathcal P_{prio}$, $\mathcal P_{sec}$ can be reduced by excluding those GTs more than $2r$ away from any GT in $\mathcal P_{prio}$, since the same MBS cannot cover two GTs that are more than $2r$ away from each other. This confines the search space to a local region near $\mathcal P_{prio}$.
For example, since the first MBS is guaranteed to cover GT 3, we can draw a dashed red circle centered at GT 3 with radius $2r$ as shown in Fig. \ref{SpiralExample}, and exclude those GTs that are outside of this circle from consideration, after which only GTs 2, 3, 4, 7, 8, and 9 are left.
This greatly reduces the problem size in (P3).
Secondly, the remaining GTs in $\mathcal P_{sec}$ are sorted in ascending order of the distance to the current MBS location $\mathbf u$, and are then successively included based on this order until they cannot be covered by the same MBS. Intuitively, the number of newly covered GTs in $\mathcal P_{sec}$ is approximately maximized.
Moreover, in step 2 of Algorithm \ref{AlgLocal}, we 
update $\mathcal P_{prio}$ ($\mathcal P_{sec}$) by including (excluding) GTs within distance $r$ to $\mathbf u$. This simple check reduces the times that the 1-center subroutine in step 3 of Algorithm \ref{AlgLocal} needs to be executed.
For example, after MBS 1 covers the boundary GTs 2, 3 and 4, the algorithm finds that GT 7 is already covered and hence does not need to call the 1-center subroutine for GT 7 subsequently.

In step 3 of Algorithm \ref{AlgLocal}, to check whether a set $\mathcal P$ of $K$ points can be covered by a single disk of radius $r$, we need to solve the 1-center problem, which finds the location $\mathbf u$ of the center from which the maximum distance to any point in $\mathcal P$ is minimized. Several algorithms exist to solve the 1-center problem, such as that in \cite{Center1n} with $O(K)$ complexity, and a more straightforward one in \cite{Center1n2} with $O(K^2)$ complexity.

For our spiral algorithm, each of the MBSs to be placed needs to run the convex hull algorithm to find the boundary GTs and list them in counterclockwise order, which has complexity $O(K\log b)$ with $b\leq K$ being the number of extreme points of the convex hull. Moreover, each MBS may also need to execute the 1-center subroutine for up to $O(K)$ times.
Since the number of placed MBSs is at most $O(K)$,
the overall computational complexity is upper-bounded by $O(K[K\log K+K\cdot C(K)])$, where $C(K)$ is the running time of the 1-center subroutine. 
Note that the actual running time could be much less than this worst-case complexity, since the size of each 1-center subroutine and the times to be executed are greatly reduced, thanks to the strategy of excluding far-away GTs and including nearby GTs in step 2 of Algorithm \ref{AlgLocal}.

\begin{figure}
\centering
\includegraphics[width=0.85\linewidth,  trim=220 0 100 30,clip]{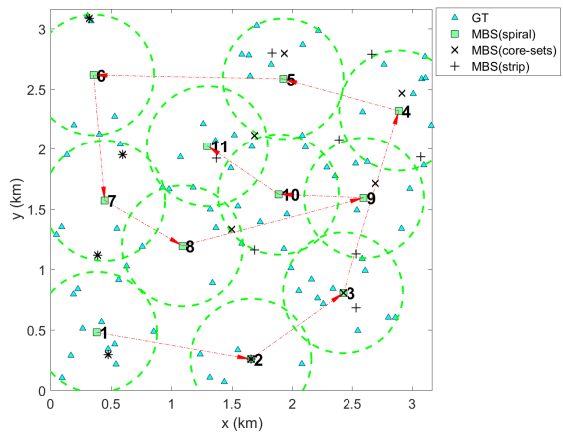} 
\caption{Solutions of the spiral, strip-based and core-sets methods to the GDC problem with 80 GTs and MBS coverage radius  $r=0.5$ km.\vspace{-3ex}} \label{GT80compare}
\end{figure}

To illustrate the final MBS placement results, we apply our spiral algorithm to a numerical example with $K=80$ GTs (denoted as triangles) randomly and independently scattered in a square region of area 10 km$^2$, where each MBS has a coverage radius $r=0.5$ km, as shown in Fig. \ref{GT80compare}.
We use dash-dotted red arrows to connect the MBSs which are successively placed along the area perimeter.
In this case, a total of 11 MBSs (denoted as green squares) are required and their connecting line looks like a spiral which starts from the area boundary and counterclockwisely revolves inwards toward the area center.

To check the optimality of our spiral algorithm, we apply the core-sets method of exponential complexity in \cite{CoresetEgypt} with stacked-depth-first branch-and-bound search to the 80 GTs' topology in Fig. \ref{GT80compare}, which yields a minimum coverage radius of 0.5231 km and 0.4829 km for 10-center and 11-center problems, respectively.
Therefore, it requires a minimum of 11 MBSs to cover all 80 GTs with a coverage radius of 0.5 km, which is the same as that achieved by our spiral algorithm. The placed MBS locations are denoted as ``$\times$" in Fig. \ref{GT80compare}. 
As a benchmark comparison, we also apply the strip-based algorithm 
in \cite{TONbackbone} to the 80 GTs' topology in Fig. \ref{GT80compare}.
It requires a total number of 13 MBSs (denoted as ``$+$" in Fig. \ref{GT80compare}) to cover all GTs, which is more than that obtained by our spiral algorithm.

\section{Numerical Results}\label{SectionNum}

In this section, we test the algorithms for two cases with $K=80$ and $K=400$ GTs, respectively.
In each case, we randomly and independently generate 5 topologies with $K$ GTs in a square region of side length $D$, and apply the algorithms to these topologies with different coverage radius $r$. For each algorithm and each $D/r$ ratio, the total number of required MBSs $M$ and the running time $t$ in seconds are averaged over the 5 topologies, respectively.
Besides the core-sets method and the strip-based algorithm, we also compare with two other heuristic schemes.
The first one is random placement, which randomly selects a location to place an MBS and removes the covered region from consideration when placing the next MBS. The process repeats until all GTs are covered. 
The second one is to run the K-means algorithm to partition the GTs to be covered by $p$ MBSs. Bisection search is performed to find the minimum number $p$ to cover all GTs.
Each of these two heuristics is executed for 100 trials on each topology and $D/r$ ratio to find the best trial with the minimum number of MBSs.
Note that the more trials of these two heuristics (hence a longer running time), the higher likelihood of finding a solution with smaller number of required MBSs.
We used the 1-center sub-routine in \cite{Center1n2} and the default initialization of the K-means function in MATLAB 2015b, which runs on Windows 10 with Intel-i5 3.5GHz PC and 8GB RAM.
The results are summarized in TABLE \ref{TableCompare}.

As observed from TABLE \ref{TableCompare}, the theoretical minimum $M_{\min}$ obtained by the core-sets method can only be found for small networks requiring only a few MBSs, e.g., $K=80$ and $M_{\min}\leq 11$ or $K=400$ and $M_{\min}\leq 8$, due to the prohibitive computational complexity of the core-sets method.
In these cases, the spiral algorithm provides the near-optimal performance in terms of $M$, but is much more time-efficient than the core-sets method.
Moreover, the spiral algorithm outperforms the strip-based algorithm in terms of $M$ while having comparable $t$ on average.
Note that the gap in $M$ between the strip-based algorithm and the spiral algorithm becomes larger as the ratio $D/r$ increases. This is expected since a larger $D/r$ ratio means more strips in the strip-based algorithm, and consequently larger performance loss. Our spiral algorithm outperforms the strip-based algorithm 
since each MBS is not restricted to cover GTs within each of the independent fixed strips, but instead can be flexibly placed to reduce outlier GTs and hence the total number of required MBSs.
Finally, the spiral algorithm also outperforms the other two heuristic schemes in terms of $M$ and $t$ on average for networks of different sizes.

\begin{table}[t]\footnotesize
\centering
\caption{Comparison between spiral algorithm and other schemes}
\addtolength{\tabcolsep}{-4pt}
\renewcommand{\arraystretch}{1.1}
\begin{tabular}{|c|c|c|c|c|c|c|c|c|c|c|c|c|}
\hline
\multicolumn{2}{|c|}{$K$} & \multicolumn{5}{c|}{80}& \multicolumn{5}{c|}{400} \\
\hline
\multicolumn{2}{|c|}{$D/r$} & 2 & 4 & 6 &8 & 10 &4 & 8 & 12 &16 & 20 \\
\hline
\multirow{2}{*}{\begin{tabular}[x]{@{}c@{}}Core-\\sets\end{tabular}} &$M$ &2.2&5.8 &10.4 &- &-  &7.8 &- &- &- & - \\
\hhline{~------------}
&$t$(s)&0.460 &5.754 &10193 &-  &- &8004 &- &- &- & - \\
\hline
\multirow{2}{*}{\begin{tabular}[x]{@{}c@{}}Spiral\end{tabular}} &$M$&2.2&5.8 &10.6 &15.4 &20.8 &8.0 &22.8 &41.6 &62.8 & 85.6\\
\hhline{~------------}
 &$t$(s)&0.116 &0.141 &0.158 &0.154 &0.151  &0.175 &0.232 &0.280 &0.300 & 0.301 \\
\hline
\multirow{2}{*}{\begin{tabular}[x]{@{}c@{}}Strip\end{tabular}} &$M$ &2.4&6.8 &12.4 &18.6 &26.8  &8.8 &25.2 &49.6 &79.8 & 111.0 \\
\hhline{~------------}
&$t$(s)&0.137 &0.130 &0.128 &0.116  &0.105 &0.338 &0.308 &0.274 &0.237 & 0.201 \\
\hline
\multirow{2}{*}{\begin{tabular}[x]{@{}c@{}}K-\\means\end{tabular}} &$M$&2.6&6.6 &11.6 &17.2 &23.0 &8.4 &26.4 &51.2 &84.4 & 120.2\\
\hhline{~------------}
 &$t$(s)&7.558 &9.151 &10.88 &11.19 &11.21  &34.13 &46.37 &61.97 &69.83 & 72.58 \\
\hline
\multirow{2}{*}{\begin{tabular}[x]{@{}c@{}}Ran-\\dom\end{tabular}} &$M$ &3.0&8.8 &17.2 &26.0 &35.2  &10.6 &36.8 &75.2 &116.6 & 162.2 \\
\hhline{~------------}
&$t$(s)&0.083 &0.329 &1.018 &1.891  &3.507 &1.246 &14.23 &39.00 &87.03 & 122.8 \\
\hline
\end{tabular}
 \label{TableCompare}
 \vspace{-1em}
\end{table}

\section{Conclusions}\label{SectionCon}
This letter proposed a new polynomial-time successive MBS placement solution for UAV-GT communications, termed as the spiral algorithm.
The proposed algorithm is compared favorably against well-known benchmark schemes in terms of the minimum number of required MBSs to cover all GTs, including the optimal core-sets based algorithm but with exponential complexity, the low-complexity strip-based algorithm, and two other heuristic schemes.
Future work could extend to the cases with additional backhaul connectivity constraint between MBSs and adaptive MBS placement subject to moving GTs.

\bibliography{IEEEabrv,BibDIRP}

\newpage

\end{document}